\documentclass[doublecol]{epl2} 

\title{Investigation of entanglement measures across the magnetization process of a highly frustrated spin-$1/2$ Heisenberg octahedral chain as a new paradigm of the localized-magnon approach}
\shorttitle{Entanglement of a highly frustrated spin-$1/2$ Heisenberg octahedral chain} 

\author{J. Stre\v{c}ka\inst{1} \and O. Krupnitska\inst{1,2} \and J. Richter\inst{3,4} }
\shortauthor{J. Stre\v{c}ka \etal}

\institute{                    
  \inst{1} Department of Theoretical Physics and Astrophysics, Faculty of Science, P.~J.~\v{S}af\'{a}rik University, Park Angelinum 9, 040 01 Ko\v{s}ice, Slovak Republic \\
	\inst{2} Institute for Condensed Matter Physics, National Academy of Sciences of Ukraine,\\ Svientsitskii Street 1, 79011 L'viv, Ukraine \\
  \inst{3} Institut f\"{u}r Physik, Otto-von-Guericke-Universit\"{a}t Magdeburg, P.O. Box 4120, 39016 Magdeburg, Germany \\
  \inst{4} Max-Planck-Institut f\"{u}r Physik komplexer Systeme, N\"{o}thnitzer Stra\ss e 38, 01187 Dresden, Germany  
}

\abstract{The bipartite entanglement across the magnetization process of a highly frustrated spin-1/2 Heisenberg octahedral chain is examined within the 
concept of localized magnons, which enables a simple calculation of the concurrence measuring a strength of the pairwise entanglement between nearest-neighbor and next-nearest-neighbor spins from square plaquettes. A full exact diagonalization of the finite-size Heisenberg octahedral chain with up to 4 unit cells (20 spins) evidences an extraordinary high precision of the localized-magnon theory in predicting measures of the bipartite entanglement at sufficiently low temperatures. While the monomer-tetramer phase emergent at low enough magnetic fields exhibits presence (absence) of the bipartite entanglement between the nearest-neighbor (next-nearest-neighbor) spins, the magnon-crystal phase emergent below the saturation field contrarily displays identical bipartite entanglement between the nearest-neighbor and next-nearest-neighbor spins. The presented results verify a new paradigm of the localized-magnon approach concerned with a simple calculation of entanglement measures.}

\begin{document}

\maketitle

\section{Introduction}

One of the most challenging tasks of modern condensed matter physics is to search solid-state resources for quantum computation and quantum information processing \cite{niel00}. Although technological implementations of some quantum algorithms such as Grover's search algorithm \cite{grov97} only requires a quantum superposition of states, several quantum algorithms as for instance Shor's factoring algorithm \cite{shor94} additionally exploit a quantum entanglement. Entanglement, as a genuine quantum phenomenon, has thus recently attracted renewed interest with regard to its wide application potential anticipated in quantum key distribution, superdense coding, quantum communication and quantum teleportation \cite{amic08,horo09}. 

It has been suggested that electron spins in solid-state systems may provide suitable platform for spin-based quantum information processing \cite{leue01,teja01,roch05,boga08,urda11}. Quantum nature of electron spins in magnetic insulators is adequately captured by the Heisenberg model \cite{matt06}. A lot of unconventional quantum phenomena and exotic quantum states of matter was revealed especially in frustrated quantum Heisenberg antiferromagnets \cite{qantum_mag,diep13,lacr11,schn20}. In spite of their complexity, magnetic properties of highly frustrated
flat-band Heisenberg spin systems may be under extreme conditions of sufficiently low temperatures and high magnetic fields satisfactorily described by a relatively 
simple concept of localized magnons originally introduced in Ref. \cite{prl02} (for comprehensive reviews see Refs.~\cite{zhit05,derz06,derz15} and references cited therein).

Although the concept of localized magnons is usually applicable just for frustrated flat-band Heisenberg spin systems at high enough magnetic fields being sufficiently close to the saturation field \cite{zhit05,derz06,derz15}, this computational method has recently proved its greater versatility when elucidating low-temperature magneto-thermodynamics of a few Heisenberg octahedral chains in a full range of magnetic fields \cite{stre17,stre18,karl19,krup20}. In the present Letter we will verify another paradigm of the theory of 
localized magnons through the calculation of entanglement measures across the overall magnetization process of a highly frustrated spin-$1/2$ Heisenberg octahedral chain originally introduced by Bose \cite{bose89,bose90,bose92}. 

\section{Model and methods}

\begin{figure}
\includegraphics[clip=on,width=\columnwidth,angle=0]{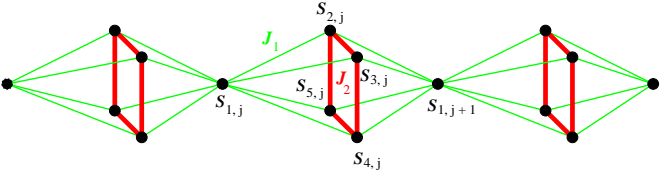}
\caption{A schematic representation of the spin-$\frac{1}{2}$ Heisenberg octahedral chain. Thick (red) lines denote the exchange interaction $J_2$ between the nearest-neighbor spins from square plaquettes, while thin (green) lines represent the exchange interaction $J_1$ between the nearest-neighbor spins from monomeric ans square-plaquette sites, respectively.}
\label{fig1}
\end{figure}

In the present work, we consider the spin-$1/2$ Heisenberg octahedral chain (see Fig.~\ref{fig1}) defined through the Hamiltonian
\begin{eqnarray}
\label{ham}
\hat{\cal H} \!\!&=&\!\! 
\sum_{j=1}^{N} \Bigl[ J_1 (\boldsymbol{\hat{S}}_{1,j} + \boldsymbol{\hat{S}}_{1,j+1}) \!\cdot\! (\boldsymbol{\hat{S}}_{2,j} + \boldsymbol{\hat{S}}_{3,j} + \boldsymbol{\hat{S}}_{4,j} + \boldsymbol{\hat{S}}_{5,j}) \Bigr.  \nonumber \\
\!\!&+&\!\! J_2 (\boldsymbol{\hat{S}}_{2,j}\!\cdot\!\boldsymbol{\hat{S}}_{3,j} + \boldsymbol{\hat{S}}_{3,j}\!\cdot\!\boldsymbol{\hat{S}}_{4,j}
+ \boldsymbol{\hat{S}}_{4,j}\!\cdot\!\boldsymbol{\hat{S}}_{5,j} + \boldsymbol{\hat{S}}_{5,j}\!\cdot\!\boldsymbol{\hat{S}}_{2,j}) \nonumber \\
\Bigl. \!\!&-&\!\! h \sum_{i=1}^{5} \hat{S}_{i,j}^{z} \Bigr],
\end{eqnarray}ö
where $\boldsymbol{\hat{S}}_{i,j} \equiv (\hat{S}_{i,j}^x, \hat{S}_{i,j}^y, \hat{S}_{i,j}^z)$ denotes a standard spin-1/2 operator assigned to a lattice site specified by two indices. The former index $i = 1-5$ determines a lattice position within the unit cell, while the latter index $j=1,\ldots,N$ determines the unit cell itself. The coupling constant $J_1>0$ denotes the antiferromagnetic exchange interaction between neighboring monomeric and square-plaquette spins, the coupling constant $J_2>0$ determines the antiferromagnetic exchange interaction between nearest-neighbor spins from the square plaquettes and the last term $h$ is the standard Zeeman's term.

It has been convincingly evidenced in Ref. \cite{stre17} that the spin-$1/2$ Heisenberg octahedral chain belongs to a valuable class of flat-band quantum Heisenberg models \cite{zhit05,derz06,derz15}, which satisfies a local conservation of the total spin angular momentum on each square plaquette. Owing to this fact, the four spins from a single square plaquette may exhibit in a highly frustrated regime $J_2/J_1 \geq 2$ just one of three available eigenstates, more specifically, the trivial fully polarized ferromagnetic eigenstate
\begin{eqnarray}
|0\rangle_j = |\!\uparrow_{2,j}\uparrow_{3,j}\uparrow_{4,j}\uparrow_{5,j}\rangle, 
\label{FM}
\end{eqnarray} 
the eigenstate with a single magnon bound to a square plaquette 
\begin{eqnarray}
|1\rangle_j = \frac{1}{2}
(|\!\downarrow_{2,j}\uparrow_{3,j}\uparrow_{4,j}\uparrow_{5,j}\rangle 
\!\!\!&-&\!\!\!|\!\uparrow_{2,j}\downarrow_{3,j}\uparrow_{4,j}\uparrow_{5,j}\rangle \nonumber \\
+ |\!\uparrow_{2,j}\uparrow_{3,j}\downarrow_{4,j}\uparrow_{5,j}\rangle
\!\!\!&-&\!\!\!|\!\uparrow_{2,j}\uparrow_{3,j}\uparrow_{4,j}\downarrow_{5,j}\rangle), \nonumber \\  
\label{MC}
\end{eqnarray}
and the singlet-tetramer eigenstate with character of a bound two-magnon state of a square plaquette
\begin{eqnarray}
|2\rangle_j = \frac{1}{\sqrt{3}}(|\!\uparrow_{2,j}\downarrow_{3,j}\uparrow_{4,j}\downarrow_{5,j}\rangle \!\!\!\!&+&\!\!\!\! |\!\downarrow_{2,j}\uparrow_{3,j}\downarrow_{4,j}\uparrow_{5,j}\!\rangle)  \nonumber \\
- \frac{1}{\sqrt{12}} (|\!\uparrow_{2,j}\uparrow_{3,j}\downarrow_{4,j}\downarrow_{5,j}\rangle \!\!\!&+&\!\!\! |\!\uparrow_{2,j}\downarrow_{3,j}\downarrow_{4,j}\uparrow_{5,j}\rangle \nonumber \\
 \hspace{-0.7cm}\!\!\!\!\!\!\!\! + |\!\downarrow_{2,j}\uparrow_{3,j}\uparrow_{4,j}\downarrow_{5,j}\rangle \!\!\!\!&+&\!\!\!\! |\downarrow_{2,j}\downarrow_{3,j}\uparrow_{4,j}\uparrow_{5,j}\rangle). 
\label{MT}
\end{eqnarray}

It has been firmly established in our previous papers \cite{stre17,stre18,karl19} that the low-temperature magneto-thermodynamics of the Heisenberg octahedral chain in the highly frustrated region $J_2/J_1 \geq 2$  can be comprehensively described within the localized-magnon theory in terms of a classical two-component lattice-gas model of hard-core monomers. The first kind of monomeric particles stands for a bound one-magnon eigenstate (\ref{MC}) with the chemical potential $\mu_1 = J_1+2J_2-h$, while the second kind of monomeric particles represents a bound two-magnon eigenstate (\ref{MT}) with the chemical potential $\mu_2=2J_1+3J_2-2h$. The chemical potentials $\mu_1$ and $\mu_2$ of both kinds of hard-core monomeric particles were assigned according to an energy difference of the localized-magnon eigenstates (\ref{MC}) and (\ref{MT}) with respect to the fully polarized ferromagnetic eigenstate (\ref{FM}) serving as reference. Bearing all this in mind, the low-energy degrees of freedom of the spin-$1/2$ Heisenberg octahedral chain can be described by a classical two-component lattice-gas model of hard-core monomers given by the following Hamiltonian
\begin{eqnarray}
{\cal H}_{\rm eff} \!= E_{\rm FM}^0 \!-\! 2hN \!-\! h\sum_{j=1}^N{S_{1,j}^z}-\mu_1 \sum_{j=1}^{N} n_{1,j} - \mu_2 \sum_{j=1}^{N} n_{2,j}, \nonumber 
\label{elm}
\end{eqnarray}
where the first term $E_{\rm FM}^0 = N (2J_1 + J_2)$ represents the zero-field energy of the fully polarized ferromagnetic state, the second term is the Zeeman's energy of the spins from the square-plaquette sites once they are all fully polarized, the third term is the Zeeman's energy of the monomeric spins and the last two terms are associated energies of both kinds of the monomeric particles as specified by their respective occupation numbers $n_{1,j}=0,1$ and $n_{2,j}=0,1$, which should not additionally violate a hard-core constraint.  
The partition function of the two-component lattice-gas model then follows from the relation
\begin{eqnarray}
{\cal Z}_{\rm eff} \!\!\!\!&=&\!\!\!\!  {\rm e}^{-\beta N\left(2J_1+J_2-2h\right)} \prod_{j=1}^N\sum_{S_{1,j}^z} {\rm e}^{\beta hS_{1,j}^z} \nonumber \\
\!\!\!\!&\times&\!\!\!\!
\sum_{n_{1,j}} \sum_{n_{2,j}} (1-n_{1,j}n_{2,j}) 
{\rm e}^{\beta (\mu_1 n_{1,j} + \mu_2 n_{2,j})}  \\
\!\!\!\!&=&\!\!\!\! {\rm e}^{-\beta N\left(2J_1+J_2-2h\right)} \! \! \left[2\cosh\left(\frac{\beta h}{2}\right) \!\! \right]^{\!N} \!\! \!\!\left(1 \!+\!{\rm e}^{\beta \mu_1} \!+\! {\rm e}^{\beta \mu_2}\right)^{\! N}\!\!\!\!, \nonumber
\end{eqnarray}
where $\beta=1/(k_{B}T)$, $k_{\rm B}$ is Boltzmann's constant, $T$ is the absolute temperature and the relevant prefactor $1-n_{1,j}n_{2,j}$ ensures 
exclusion of double occupancy of one and same site by two different kinds of monomeric particles from consideration. The associated Helmholtz free energy per elementary unit can be calculated according to the formula
\begin{eqnarray}
F_{\rm eff}  \!\!&=&\!\! -k_{\rm B} T \lim_{N \to \infty} \frac{1}{N} \ln {\cal Z}_{\rm eff} \nonumber \\
\!\!&=&\!\! 2J_1 \!+\! J_2 \!-\! 2h\!- k_{\rm{B}}T\ln\left[2 \cosh\left(\frac{\beta h}{2}\right)\right] \nonumber \\
\!\!&-&\!\! \!k_{\rm{B}}T\ln\left(1 + {\rm e}^{\beta \mu_1} \!+\! {\rm e}^{\beta \mu_2} \right).
\label{lmgfe}
\end{eqnarray}
The total magnetization per unit cell can be obtained from Eq. (\ref{lmgfe}) using the standard relation
\begin{eqnarray}
M = -\frac{\partial F_{\rm eff}}{\partial h} = 2 + \frac{1}{2} \tanh\left(\frac{\beta h}{2}\right) - \frac{{\rm e}^{\beta \mu_1} + 2{\rm e}^{\beta \mu_2}}{1 + {\rm e}^{\beta \mu_1} + {\rm e}^{\beta \mu_2}}.
\label{M}
\end{eqnarray}

The main goal of the present work is to investigate the strength of the bipartite entanglement between the nearest- and next-nearest-neighbor spins from square plaquettes of the spin-$1/2$ Heisenberg octahedral chain. To this end, one may take advantage of the quantity concurrence originally introduced by Wootters \cite{woot98}, which may be used as a suitable measure of the bipartite entanglement for any two-qubit system. It has been found by Amico and coworkers \cite{amic04} that the concurrence can be alternatively calculated from the local observables such as pair correlation functions and magnetization, which can be rather easily calculated also within the concept of localized magnons. Accordingly, the concurrence $C_{nn}$ and $C_{nnn}$ quantifying the degree of the bipartite entanglement between the nearest-neighbor and next-nearest-neighbor spins from the square plaquettes follows from the relations  
\begin{eqnarray}
\label{conc_nn}
C_{nn}\!\!\!&=&\!\!\! {\rm max} \Bigg\{0, 4|\langle {\hat S}^{x}_{2,j} {\hat S}^{x}_{3,j}\rangle| \\
\!\!\!&-&\!\!\! 2\sqrt{\left(\frac{1}{4}+\langle {\hat S}^{z}_{2,j} {\hat S}^{z}_{3,j}\rangle\right)^2-\left(\frac{1}{2}\langle {\hat S}^{z}_{2,j} + {\hat S}^{z}_{3,j}\rangle \right)^2}\Bigg\} \nonumber
\end{eqnarray}
and 
\begin{eqnarray}
\label{conc_nnn}
C_{nnn}\!\!\!&=&\!\!\!{\rm max}\Bigg\{0, 4|\langle {\hat S}^{x}_{2,j} {\hat S}^{x}_{4,j}\rangle| \\
\!\!\!&-&\!\!\! 2\sqrt{\left(\frac{1}{4}+\langle{\hat S}^{z}_{2,j} {\hat S}^{z}_{4,j}\rangle\right)^2-\left(\frac{1}{2}\langle {\hat S}^{z}_{2,j} + {\hat S}^{z}_{4,j}\rangle \right)^2}\Bigg\}. \nonumber
\end{eqnarray}
The pair correlation functions entering into Eqs. (\ref{conc_nn}) and (\ref{conc_nnn}) can be readily calculated within the presented localized-magnon theory according to the formulas
\begin{eqnarray}
\langle {\hat S}^{\alpha}_{2,j} {\hat S}^{\alpha}_{3,j}\rangle \!\!\!\!&=&\!\!\!\! \frac{1}{{\cal Z}_{\rm eff}} \!\! \sum_{\{\! S_{1,j}^z \!\}} \sum_{\{\! n_{1,j} \!\}} \sum_{\{\! n_{2,j}\!\}} \sum_{k=0}^2
\langle k | {\hat S}^{\alpha}_{2,j} {\hat S}^{\alpha}_{3,j} | k \rangle_{\!j} \, {\rm e}^{-\beta {\cal H}_{\rm eff}} \nonumber \\
\!\!\!\!&=&\!\!\!\!
\frac{C^{\alpha\alpha}_{kk}+C^{\alpha \alpha}_{kk}{\rm e}^{\beta \mu_1}+C^{\alpha\alpha}_{kk}{\rm e}^{\beta \mu_2}}{1 \!+\!{\rm e}^{\beta \mu_1} \!+\! {\rm e}^{\beta \mu_2}}
\label{c_nn}
\end{eqnarray}
and
\begin{eqnarray}
\langle {\hat S}^{\alpha}_{2,j} {\hat S}^{\alpha}_{4,j}\rangle \!\!\!\!&=&\!\!\!\! \frac{1}{{\cal Z}_{\rm eff}} \!\! \sum_{\{\! S_{1,j}^z \!\}} \sum_{\{\! n_{1,j} \!\}} \sum_{\{\! n_{2,j}\!\}} \sum_{k=0}^2
\langle k | {\hat S}^{\alpha}_{2,j} {\hat S}^{\alpha}_{4,j} | k \rangle_{\!j} \, {\rm e}^{-\beta {\cal H}_{\rm eff}} \nonumber \\
\!\!\!\!&=&\!\!\!\! 
\frac{K^{\alpha\alpha}_{kk}+K^{\alpha \alpha}_{kk}{\rm e}^{\beta \mu_1}+K^{\alpha\alpha}_{kk}{\rm e}^{\beta \mu_2}}{1 \!+\!{\rm e}^{\beta \mu_1} \!+\! {\rm e}^{\beta \mu_2}},
\label{c_nnn}
\end{eqnarray} 
which are defined through three different mean values $C^{\alpha \alpha}_{kk} = \langle k| {\hat S}^{\alpha}_{2,j} {\hat S}^{\alpha}_{3,j}|k\rangle_j$ and $K^{\alpha \alpha}_{kk} = \langle k| {\hat S}^{\alpha}_{2,j} {\hat S}^{\alpha}_{4,j}|k\rangle_j$ $(k = 0, 1, 2; \alpha = x, z)$ of the respective spin operators corresponding to the three available eigenstates (\ref{FM})-(\ref{MT}) of the $j$-th square plaquette. The local magnetizations $m_{23} = \frac{1}{2}\langle {\hat S}^{z}_{2,j} + {\hat S}^{z}_{3,j}\rangle$ and $m_{24} = \frac{1}{2}\langle {\hat S}^{z}_{2,j} + {\hat S}^{z}_{4,j}\rangle$ can be calculated in analogous way as presented for the pair correlation functions, but they both are formally identical with the single-site magnetization $m_{23} = m_{24} = M/5$ as obtained from Eq. (\ref{M}) for the magnetization of the unit cell.

To provide an independent check of the reliability of the localized-magnon theory in predicting measures of the bipartite entanglement in the highly frustrated parameter space $J_2/J_1 \geq 2$ we have additionally performed the full exact diagonalization (ED) for the spin-1/2 Heisenberg octahedral chain up to 20 spins ($N$ = 4 unit cells) imposing periodic boundary conditions using  J\"org Schulenburg's  {\it spinpack} \cite{schu10,rich10}. The main advantage of the ED method is to provide the direct access to all local pair correlation functions and local magnetizations, which are needed according to Eqs. (\ref{conc_nn}) and (\ref{conc_nnn}) for the calculation of the concurrence for the nearest-neighbor and next-nearest-neighbor spin pairs, respectively. It should be emphasized that the numerical ED of the finite-size spin-1/2 Heisenberg octahedral chain in fact provides a rigorous check of the developed localized-magnon approach, because the predicted results for a classical one-dimensional lattice-gas model defined through the Hamiltonian (\ref{elm}) falls into a monomer universality class for which the results are completely independent of the system size \cite{derz06,derz15}.

\section{Results and discussion}

In this part we will comprehensively examine the strength of the bipartite entanglement between the nearest- and next-nearest-neighbor spin pairs from square plaquettes along the magnetization process of the highly frustrated spin-1/2 Heisenberg octahedral chain. Before doing this, it is worthwhile to recall that the spin-1/2 Heisenberg octahedral chain exhibits in a highly frustrated regime $J_2/J_1 \geq 2$ just three different ground states as convincingly evidenced in Ref. \cite{stre17}. At sufficiently low magnetic fields $h < J_1 + J_2$ one detects as a respective ground state the monomer-tetramer (MT) phase with a singlet-tetramer state of all square plaquettes
\begin{eqnarray}
|{\rm MT} \rangle \!\!=\!\! \prod_{j=1}^N \! |\!\uparrow_{1,j}\rangle \!\otimes\! 
\frac{1}{\sqrt{3}}\Bigl[\!(|\!\uparrow_{2,j}\downarrow_{3,j}\uparrow_{4,j}\downarrow_{5,j}\rangle \!\!\!\!\!\!&+&\!\!\!\!\!\! |\!\downarrow_{2,j}\uparrow_{3,j}\downarrow_{4,j}\uparrow_{5,j}\rangle)  \nonumber \\
- \frac{1}{2} (|\!\uparrow_{2,j}\uparrow_{3,j}\downarrow_{4,j}\downarrow_{5,j}\rangle \!\!\!\!\!\!&+&\!\!\!\!\!\! |\!\uparrow_{2,j}\downarrow_{3,j}\downarrow_{4,j}\uparrow_{5,j}\rangle \nonumber \\
+ |\!\downarrow_{2,j}\uparrow_{3,j}\uparrow_{4,j}\downarrow_{5,j}\rangle \!\!\!\!\!\!&+&\!\!\!\!\!\! |\!\downarrow_{2,j}\downarrow_{3,j}\uparrow_{4,j}\uparrow_{5,j}\rangle) \Bigr]\!, \nonumber \\
\label{MTP}
\end{eqnarray}
the magnon-crystal (MC) phase with a single magnon bound to each square plaquette is the respective ground state at moderate magnetic fields $h \in (J_1 + J_2, J_1 + 2J_2)$ 
\begin{eqnarray}
|{\rm MC}\rangle \!\!=\!\! \prod_{j=1}^N \! |\!\uparrow_{1,j}\rangle \!\otimes\! \frac{1}{2}
(|\!\downarrow_{2,j}\uparrow_{3,j}\uparrow_{4,j}\uparrow_{5,j}\rangle 
\!\!\!\!\!\!&-&\!\!\!\!\!\!|\!\uparrow_{2,j}\downarrow_{3,j}\uparrow_{4,j}\uparrow_{5,j}\rangle \nonumber \\
+|\!\uparrow_{2,j}\uparrow_{3,j}\downarrow_{4,j}\uparrow_{5,j}\rangle
\!\!\!\!\!\!&-&\!\!\!\!\!\!|\!\uparrow_{2,j}\uparrow_{3,j}\uparrow_{4,j}\downarrow_{5,j}\rangle), \nonumber \\  
\label{MCP}
\end{eqnarray}
and finally, the trivial fully polarized ferromagnetic (FM) phase emerges above the saturation field $h > J_1 + 2J_2$
\begin{eqnarray}
|{\rm FM} \rangle = \prod_{j=1}^N \! |\!\uparrow_{1,j}\uparrow_{2,j}\uparrow_{3,j}\uparrow_{4,j}\uparrow_{5,j}\rangle. 
\label{FMP}
\end{eqnarray} 

\begin{figure}
\includegraphics[clip=on,width=\columnwidth,angle=0]{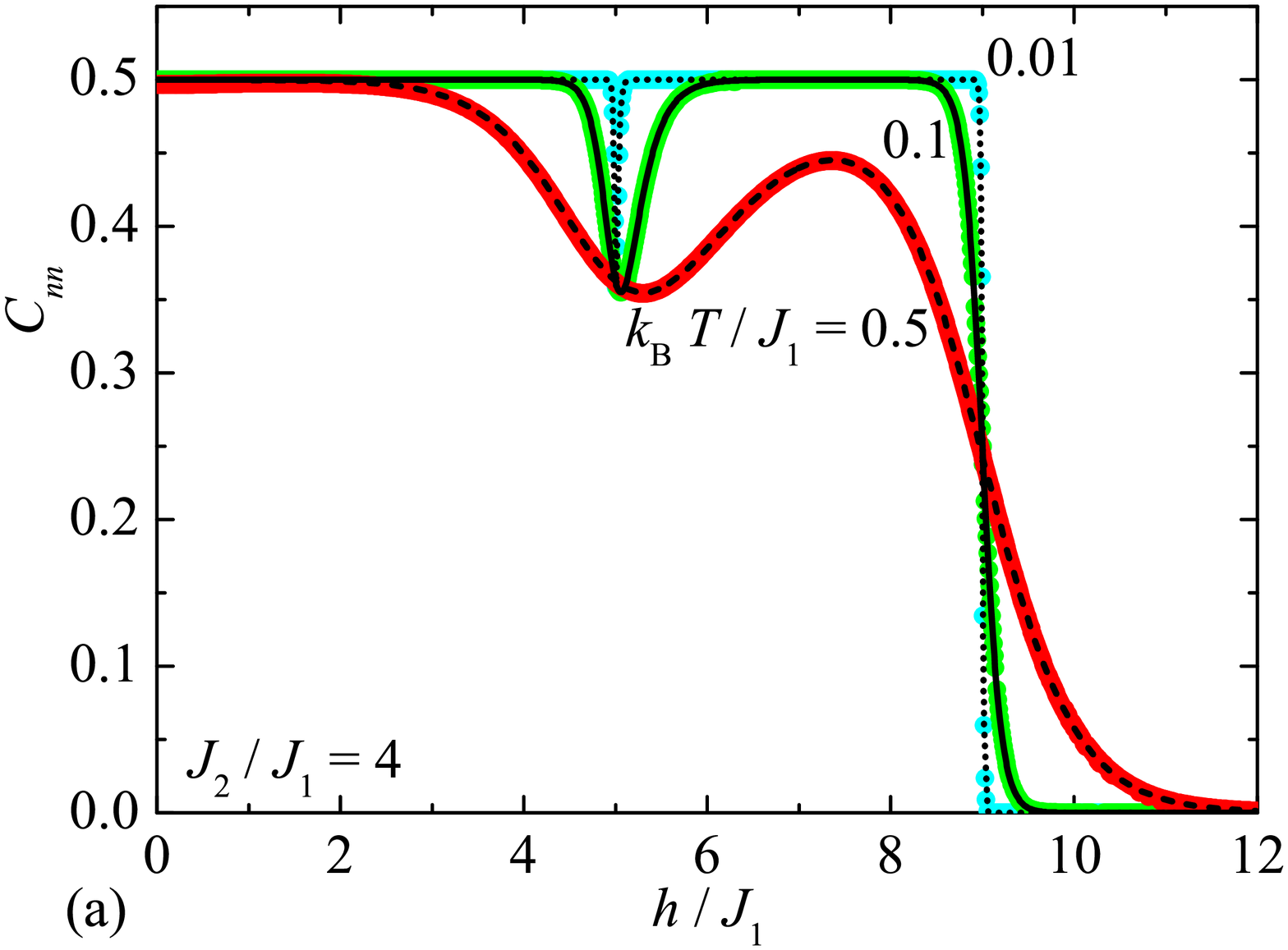}
\includegraphics[clip=on,width=\columnwidth,angle=0]{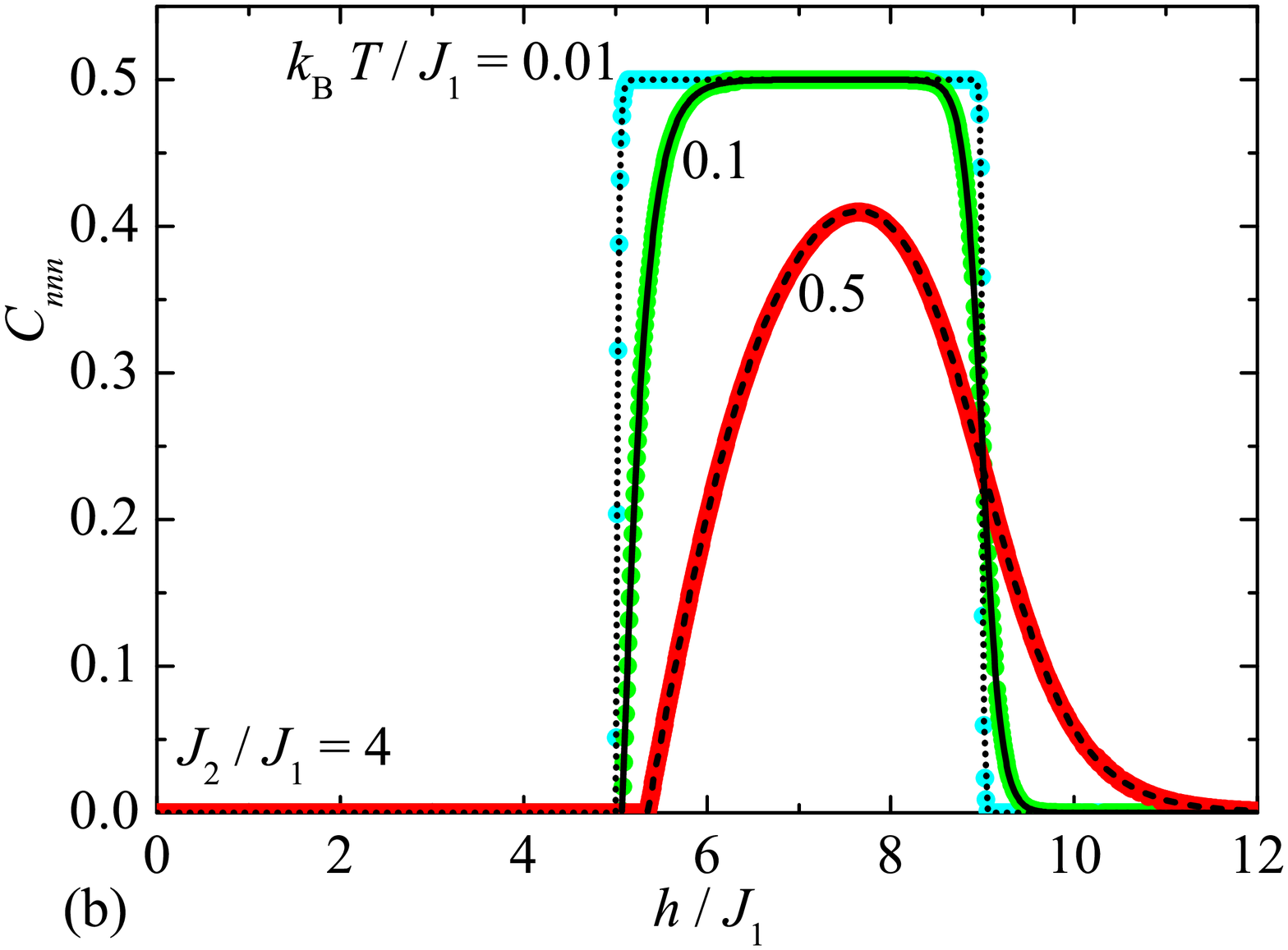}
\vspace*{-1cm}
\caption{Isothermal dependencies of the quantum concurrence on a magnetic field, which quantifies the bipartite entanglement between the nearest-neighbor $C_{nn}$ (a) and next-nearest-neighbor $C_{nnn}$ (b) spin pairs within square plaquettes. Thin black lines were calculated from Eqs. (\ref{conc_nn})-(\ref{c_nnn}) within the concept of localized magnons, while thick color symbols were obtained from a full ED of the finite-size octahedral chain with 4 unit cells (20 spins).}
\label{fig2}
\end{figure}

At first, let us compare results for the concurrence, which serves as a measure of the bipartite entanglement, obtained from the concept of localized magnons with that ones obtained from
the full ED method for the finite-size octahedral spin chain with 4 unit cells (20 spins). To this end, isothermal dependencies of the concurrence on a magnetic field are depicted in Fig. \ref{fig2}(a) and (b) at three different temperatures for the nearest-neighbor and next-nearest-neighbor spin pairs of square plaquettes, respectively. It is worthy to recall that the formula derived within the concept of localized magnons is independent of the system size and hence, it can be directly compared with the finite-size results stemming from the ED method. 

It is obvious from Fig. \ref{fig2}(a) and (b) that the results derived by making use of the localized-magnon method perfectly coincide with the ED data up to moderate temperatures $k_{\rm B} T/J_1 \approx 0.5$. This observation would suggest that the localized-magnon approach provides a suitable tool for investigation of entanglement measures at low up to moderate temperatures. Besides, it follows from Fig. \ref{fig2}(a) that the bipartite entanglement between the nearest-neighbor spin pairs of square plaquettes is equally strong in the MT and MC ground states, whereas the relevant field-driven quantum phase transition between them is manifested at low enough temperatures as a sharp narrow minimum of the concurrence that becomes rounder and wider upon increasing of temperature. A sharp stepwise drop of the concurrence $C_{nn}$ calculated for the nearest-neighbor spin pairs, which appears at sufficiently low temperatures in a vicinity of the saturation field, notably changes to a more gradual decline forming an extended high-field tail observable at higher temperatures. The nonzero concurrence between the nearest-neighbor spin pairs above the saturation field implies a substantial role of low-lying thermal excitations from the classical FM ground state towards the entangled MC state.  

The results presented in Fig. \ref{fig2}(b) indicate a complete absence of the bipartite entanglement between the next-nearest-neighbor spin pairs of the square plaquettes at low enough magnetic fields and this statement holds true irrespective of temperature. However, the concurrence $C_{nnn}$ for the next-nearest-neighbor spin pairs displays at low enough temperatures a peculiar rectangular-shaped field dependence with a sharp rise and fall at lower and upper critical fields of the MC ground state. It could be thus concluded that the pairwise entanglement between the next-nearest-neighbor spin pairs is at low enough temperatures fully absent in the low-field range inherent to the MT ground state with two magnons bound to each square plaquette, then it is provisionally strengthened in the range of moderate magnetic fields corresponding to the MC ground state with a single magnon bound to each square plaquette before it finally breaks down at sufficiently high magnetic fields. Similar findings have been recently reported also for the spin-1/2 Heisenberg distorted tetrahedron \cite{karl20}. Note furthermore that the sharp rectangular-shaped field dependence of the concurrence for the next-nearest-neighbor spin pairs is gradually smeared out upon increasing temperature quite similarly as the concurrence for the nearest-neighbor spin pairs does. 

\begin{figure}
\includegraphics[width=\columnwidth]{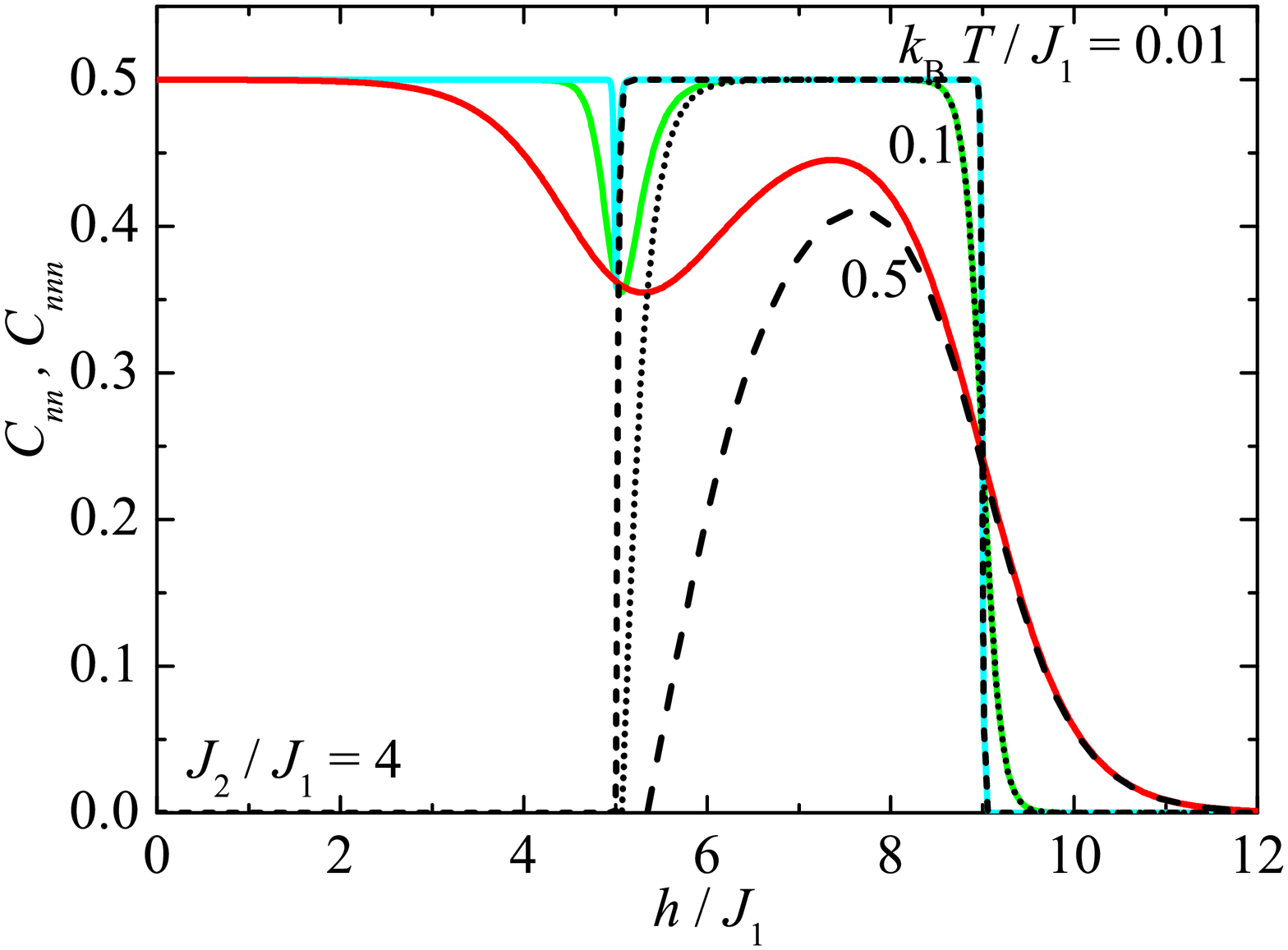}
\vspace*{-1cm}
\caption{A comparison between the concurrences $C_{nn}$ and $C_{nnn}$, which serve as a measure of the bipartite entanglement between the nearest-neighbor (color solid lines) and next-nearest-neighbor (black broken lines) spins within square plaquettes, calculated from Eqs. (\ref{conc_nn})-(\ref{c_nnn}) within the localized-magnon theory.} 
\label{fig3}
\end{figure}

Last but not least, we will directly compare the strength of the bipartite entanglement between the nearest-neighbor and next-nearest-neighbor spins quantified by means of the respective values of the concurrence $C_{nn}$ and $C_{nnn}$, which are displayed together in Fig. \ref{fig3} as a function of the magnetic field at three selected temperatures. It surprisingly turns out that the measures of the bipartite entanglement between the nearest-neighbor and next-nearest-neighbor spin pairs are not only qualitatively but also quantitatively the same in a certain range of magnetic fields, which crucially depends on temperature. As a matter of fact, both concurrences become identical at low enough temperatures for any magnetic field that slightly exceeds the lower critical field, while the rise of temperature generally shifts this peculiar coincidence to higher magnetic fields though its onset never exceeds the upper critical field. A gradual tail-like decay of the concurrence, which can be observed above the saturation field due to thermal activation of the excited MC state, can be accordingly characterized by a mutual interplay of alike pairwise entanglement between the nearest-neighbor and next-nearest-neighbor spins.

\section{Conclusions}

In the present Letter we have investigated in detail the bipartite entanglement between the nearest-neighbor and next-nearest-neighbor spin pairs of a highly frustrated spin-1/2 Heisenberg octahedral chain in the presence of external magnetic field. It has been demonstrated that the concept of localized magnons can be straightforwardly adapted for a  calculation of the quantity concurrence, which may serve as a measure of the pairwise entanglement between nearest-neighbor and next-nearest-neighbor spins. To provide an independent check of the localized-magnon method we have performed a full ED of the finite-size spin-1/2 Heisenberg octahedral chain with 4 unit cells (20 spins), which has convincingly evidenced an extraordinary high precision of the concept of localized magnons in predicting measures of the bipartite entanglement at sufficiently low temperatures. The developed concept thus sheds light on a unprecedented feature of the localized-magnon theory, which can be straightforwardly adapted in order to calculate the respective entanglement measures for a wide class of flat-band quantum Heisenberg spin models from different universality classes comprehensively reviewed in Refs. \cite{zhit05,derz06,derz15}. 

As far as the highly frustrated spin-1/2 Heisenberg octahedral chain as a paradigmatic example of flat-band quantum spin model with extremely strong quantum correlations is concerned, 
we have provided evidence that the MT phase emergent at sufficiently low magnetic fields exhibits presence (absence) of bipartite entanglement between the nearest-neighbor (next-nearest-neighbor) spins, while the MC phase emergent at moderate magnetic fields contrarily displays identical bipartite entanglement between the nearest-neighbor and next-nearest-neighbor spins. It could be thus concluded that the rising magnetic field may provisionally reinforce a pairwise entanglement between the next-nearest-neighbor spins when still preserving the same strength of quantum correlations between the nearest-neighbor spins in the range of moderately strong magnetic fields. 

\acknowledgments
This work was financially supported by Slovak Research and Development Agency provided under the contract No. APVV-16-0186 and by The Ministry of Education, Science, Research and Sport of the Slovak Republic provided under the grant No. VEGA 1/0105/20. J.R. thanks the DFG for financial support (grant RI 615/25-1).

\end{document}